\documentclass[prl,twocolumn,showpacs,preprintnumbers,amsmath,amssymb]{revtex4}

\usepackage{graphicx}
\usepackage{dcolumn}
\usepackage{bm}
\usepackage{color}
\usepackage{ulem}

\begin{document}

\preprint{APS/123-QED}

\title{Decoherence via induced dipole collisions in an ultracold gas}

\author{Anthony R. Gorges, Mathew S. Hamilton, and Jacob L. Roberts}
 
\affiliation{%
Department of Physics, Colorado State University, Fort Collins, CO 80523
}%

\date{\today}

\begin{abstract}
We have studied the effects of loading $^{87}$Rb into a far off resonant trap (FORT) in the presence of an ultracold cloud of $^{85}$Rb. The presence of the $^{85}$Rb resulted in a marked decrease of the $^{87}$Rb load rate. This decrease is consistent with a decrease in the laser cooling efficiency needed for effective loading. While many dynamics which disrupt loading efficency arise when cooling in a dense cloud of atoms (reabsorption, adverse optical pumping, etc.), the large detuning between the transitions of $^{85}$Rb and $^{87}$Rb should isolate the isotopes from these effects. For our optical molasses conditions we calculate that our cooling efficiencies require induced ground-state coherences. We present data and estimates which are consistent with heteronuclear long-ranged induced dipole-dipole collisions disrupting these ground state coherences, leading to a loss of optical trap loading efficiency.
\end{abstract}

\maketitle
Anisotropic interactions between ultracold atoms can be created by inducing electric dipole moments with light.  Such induced dipoles are responsible for light-assisted collisions \cite{Weiner1999}, allow the formation of long-range molecules \cite{Jones2006, Stwalley1978}, and have been discussed in the context of quantum computation \cite{Brennen2000, Petrosyan} and studies of anisotropic interactions in Bose-Einstein condensates \cite{Gio2002, Mazets}.  At sufficiently close internuclear separation ($R$), these dipoles have a $\frac{1}{R^3}$ interaction potential. Since the ground state van der Waals forces scale as $\frac{1}{R^6}$, the induced dipoles can dominate these interactions at relatively long ranges. Because these dipoles are anisotropic, the interaction potential is dependent on atomic spin polarization.  In this Letter, we show that such spin-dependent long-range dipole-dipole collisions between $^{85}$Rb and $^{87}$Rb atoms can lead to decoherence in the atoms' ground states during laser cooling. Collision-induced decohering effects have been reported in other systems, although the dynamics differ from system to system \cite{Tannoudji1992, Lisdat2009, Hornberger2003}. We believe these decohering collisions explain our observation of a reduced optical trap loading rate for $^{87}$Rb atoms in the presence of a diffuse gas of $^{85}$Rb atoms.

Our sensitivity to decoherence is due to the relatively large detuning (120 MHz to the red of the cycling transition in $^{87}$Rb) that optimized the number of atoms loaded into our shallow ($\sim$120 $\mu$K) Far-off resonant optical trap (FORT) \cite{Hamilton2009}. At this large detuning, sub-Doppler cooling (e.g.\ polarization gradient cooling) is especially dependent upon induced coherences in the ground state m$_{F}$ magnetic sublevels which facilitate coherent multi-photon processes that produce ground-state m$_{F}$ population transfer and velocity-dependent scattering \cite{Dalibard, Minogin}. A relationship needs to be maintained between the light polarization and the atom polarization in order for this cooling to be effective. For our experimental conditions, the incoherent photon scattering rate from a single pair of beams is $\frac{1}{24\mu s}$, insufficient to maintain an atom/light field polarization correlation. The atoms move, on average, 2400 nm in that time, while the light field polarizations pattern varies on a scale of $\sim$200 nm. Thus the majority of the damping force in our optical molasses is due to coherent processes.

It is known both theoretically \cite{Hillenbrand1994, Hillenbrand1995, Ellinger1997} and experimentally \cite{Grego1996, Drewsen1994, Voroscovs2005} that the performance of sub-Doppler cooling decreases as the density of the gas being cooled increases.  However, such a disruption via light scattering in not expected from the introduction to the cloud of a second type of atom with different resonant transition frequencies.  For instance, in the light field of our optical molasses used to cool $^{87}$Rb, any $^{85}$Rb atoms would scatter photons at a rate $\sim$400 times slower than the $^{87}$Rb itself. Nevertheless, in the course of studying the loading of $^{87}$Rb into a shallow FORT we found evidence for a decrease in laser cooling performance due to the presence of a diffuse cloud of $^{85}$Rb atoms.  While light scattering cannot be responsible for this decrease, the density of the $^{85}$Rb cloud also means any collisions would have to be much longer ranged than implied by the cross sections of either measured inelastic light-assisted collisions or usual (i.e. no light present) elastic collisions.  Calculations show that dipoles induced by the laser light are big enough to produce a sufficiently long-ranged interaction.  

To characterize this laser cooling disruption we measured the load rate of $^{87}$Rb into the FORT both with and without the presence of $^{85}$Rb atoms. Since the FORT is essentially a conservative potential, the load rate is a function of how well atoms entering the FORT region can be laser cooled into the optical potential and become trapped.  The FORT is loaded directly from a Magneto Optic Trap (MOT) which uses a cooling laser (tuned near the $^{87}$Rb D2 cycling transition) and a repump laser (on the $^{87}$Rb D1 line). The load rate was determined by measuring the number of $^{87}$Rb atoms trapped in the FORT as a function of time after the FORT was turned on. The procedure for measuring the $^{87}$Rb load rate with no $^{85}$Rb atoms (``alone'') was very similar to that used to measure the $^{87}$Rb load rate with $^{85}$Rb atoms present (``dual''). For the alone loading, $^{87}$Rb was first collected into its MOT. To prepare for a FORT load, the hyperfine repump power was decreased and the cooling laser detuning was increased to 120 MHz to the red of the cycling transition. The FORT was then turned on, which was accomplished via an acusto-optical-modulator (AOM). Once the FORT was turned on, the atoms began to load and the trap was allowed to load for a set time (evolution time). After the desired evolution time was reached, the $^{87}$Rb trapping light was shut off and the trapped atoms were held in the FORT for 100 ms to allow any untrapped atoms time to fall away before imaging. At the end of this hold time the atoms were released from the FORT and were then imaged using standard absorption imaging techniques. 

For the dual data, the sequence was mostly the same as above. Here however, $^{85}$Rb was collected in a MOT which spatially overlaped the $^{87}$Rb MOT. Before the FORT was turned on, one of the $^{85}$Rb MOT lasers was turned off. This prevented the $^{85}$Rb from loading significantly into the FORT. The evolution time, additional hold, and imagining was all the same as in the alone sequence. The results of these measurements are shown in Figure 1.

Once we had measured the number of trapped $^{87}$Rb atoms as a function of evoution time, we then extracted a loading rate ($R$). The number of $^{87}$Rb atoms trapped as a function of time is given by
\begin{eqnarray}
\dot{N}_{87}=-\beta N^{2}_{87} - \beta ' N_{87}N_{85} + R
\end{eqnarray}
where $\beta$ and $\beta '$ are effective loss rates for the homonuclear and heteronuclear losses respectively, $N_{85}$ and $N_{87}$ are the $^{85}$Rb and $^{87}$Rb number in the FORT respectively, and $R$ is the load rate into the FORT. 

$\beta$ is measured through the study of homonuclear FORT loads following the methodology in \cite{Hamilton2009}. $\beta$' was measured in a series of separate experiments. Both $^{85}$Rb and $^{87}$Rb were loaded into the trap.  Then, one of the $^{85}$Rb trapping lasers is turned off, preventing additional $^{85}$Rb loading.  The number of both isotopes was then measured as a function of time. With the load cut off, the number of $^{85}$Rb atoms decreased due to heteronuclear and off-resonant homonuclear light-assisted collisions. The value of $\beta$' was then determined from this loss data. 

Once $\beta$ and $\beta$' were known, all data was then fit to Equation (1) to determine R. The load rate reduction caused by the presence of the non-resonant isotope is apparent in Figure 1. The presence of the $^{85}$Rb thus causes a 27$\pm$5$\%$ decrease in the overall loading rate.  This difference was not sensitive to the precise value of $\beta$ and $\beta$'. Variations of the loss coefficients of 50\% or more produced no noticeable effect on the ratio of R$_{alone}$/R$_{dual}$. While the $^{85}$Rb was not actively loaded into the FORT, there were always some amount loaded by turning on the FORT \cite{Wu2006}. There were also $^{85}$Rb atoms which passed through the FORT volume which were not trapped but contributed to the $^{85}$Rb density in the trap region. The number of $^{85}$Rb was low (0.4x10$^{6}$) compared to the number of $^{87}$Rb ultimately loaded into the FORT.

\begin{figure}
\includegraphics{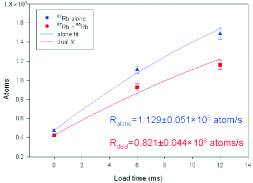}
\caption{\label{label} (color online) FORT loading data for $^{87}$Rb alone and $^{87}$Rb with $^{85}$Rb present. The plot shows number of $^{87}$Rb atoms in the FORT vs. evolution time. Triangle (blue) data are $^{87}$Rb alone, while the square (red) data are $^{87}$Rb in the presence of an $^{85}$Rb MOT. $^{85}$Rb was purposefully loaded only minimally into the FORT so as not to produce a measurable effect via light assisted collisional loss.}
\end{figure}

To ensure that what was being observed was an actual load rate disruption into the FORT, several checks were performed. First, the number of the $^{87}$Rb MOT was measured with and without the presence of $^{85}$Rb. Less than a 10\% reduction in number was measured. Such a small reduction should not significantly impact $^{87}$Rb loading as $R$ is expected to scale as $N^{1/3}$ (where $N$ is the number of atoms in the MOT) \cite{density}, which we confirmed experimentally. Second, we checked to see if the load rate reduction was due to the presence of the $^{85}$Rb MOT light itself. To test this, we loaded $^{87}$Rb atoms alone with and without one or both of $^{85}$Rb's lasers on. For the case where both lasers were on, they were turned on just before the FORT was turned on, preventing $^{85}$Rb atoms from being trapped in a MOT prior to FORT loading. In all these cases, the measured load rate showed no significant reduction due to the $^{85}$Rb trap light. In fact, the $^{87}$Rb alone data displayed in Fig.\ 1 were collected in the full $^{85}$Rb trap light field, but with the light turned on too late to collect any $^{85}$Rb into a MOT. 

\begin{figure}
\includegraphics{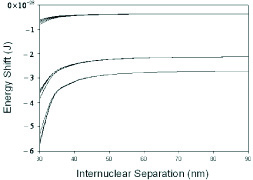}
\caption{\label{label} The dressed ground state potentials for induced dipole-dipole interaction between $^{87}$Rb F=2 ground state and $^{85}$Rb F=2 ground state. The potentials have significant slope even at long range ($\sim$50 nm) and are state-dependent.} 
\end{figure}

In principle, an increase in evaporation during the 100 ms hold time or light-assisted inelastic collisions could be responsible for the observed reduction in trapped atom number during the loading. Measurements indicated that the increase in evaporation was negligible.  With respect to the light-assisted collisions, we included them explicitly in our determination of $R$. One of the characteristics of these collisions is that they are strongly state-dependent \cite{Gorges}. Thus we investigated the load rate reduction when the $^{85}$Rb were all in the lower hyperfine state and when they were all in the upper hyperfine state. We found no observable dependence on the $^{87}$Rb load rate reduction due to $^{85}$Rb hyperfine state. 

Given the small expected effect from $^{85}$Rb light scattering, we calculated the collision rate necessary to produce the observed reduction in $^{87}$Rb loading given the low density of $^{85}$Rb atoms present.  This calculation indicated that such a collision would have to have an effective range of $\sim$50 nm.  From our observations this collision would also have to be dependent on the presence of the $^{87}$Rb cooling light.  Both of these requirements are met by considering the dipoles induced in $^{87}$Rb and $^{85}$Rb by the $^{87}$Rb cooling light.  Figure 2 shows dressed ground state $^{87}$Rb/$^{85}$Rb interatomic potentials including these induced dipole-dipole interactions. The resulting forces can be large; at an internuclear separation of 50 nm, an atom pair experiences an acceleration of $\sim$20,000 m/s$^{2}$ for the steepest potential shown. Furthermore, these forces are state-dependent.

\begin{figure}
\includegraphics{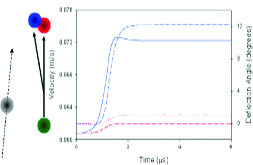}
\caption{\label{label} (color online) A classical view of the wavepacket splitting. As an $^{87}$Rb wavepacket (green) experiences the long range dipole-dipole interaction with an $^{85}$Rb wavepacket (gray), it is split into two by the state dependent forces (red and blue). The graph shows the different velocities and deflection angles experienced by the split wavepackets. High deflection is depicted by the solid line (velocity) and dashed line (angle). Low deflection is depicted by the dash-dot line (velocity) and the dotted line (angle)} 
\end{figure}

As in a Stern-Gerlach type experiment, state-dependent forces lead to a wavepacket splitting. Any state-dependent wavepacket splitting will tend to disrupt the existing ground-state coherences in several different ways.  The split wavepackets follow different paths through the cooling light field polarization field resulting in different forces, diffusion coefficients, and equilibrium m$_F$ state distributions, an example of which is shown in Fig.\ 3. In the lab frame, these collisions can induce significant changes in the wavepacket average kinetic energy, which also leads to a sudden change in forces, diffusion coefficient, and equilibrium state for the different wavepackets. Also, the optical molasses closely couples momentum states separated by 2$\hbar k$ and a splitting will suddenly shift the phase of those relationships.  As the split wavepackets propagate with different speeds and directions, additional phase shifts develop.  Additionally, similar in magnitude and range state-dependent forces are produced in collisions with atoms in the $^{87}$Rb F=1 state due to the repump light. While that light is weaker than the cooling light, it is much closer to resonance ($\sim$ 8 MHz blue detuned).

To estimate the effect of these collision more quantitatively we created a model that used a combination of dressed state potentials and classical atom trajectories to calculate the amount of state-dependent deflection that occurred in a collision.  A collision was labeled ``disruptive'' if the maximum deflection difference between different states was greater than $\hbar k$ in the lab frame, where $k$ is the wave number associated with the cooling laser.  This threshold was chosen because it indicated a significant amount of state-dependent deflection ($>$0.1 rad) and average kinetic energy change ($>$20\%), and led to significant phase shifts for the different wavepackets (modeling their propagation as being in free space).  In any case, the results are not strongly dependent on the choice of this threshold due to the steepness of the potentials.  By performing a thermal average over collision energies and impact parameters our model indicated that a 25\% load rate reduction was reasonable if one disruptive collision during a transit of the optical trap region prevented an $^{87}$Rb atom from being loaded when it otherwise would have been.  Optical Bloch equations for the internal states showed that the collision-induced optical pumping to the F=1 ground state due to loss of coherence, diffusive heating, and loss of velocity damping were sufficient so that a single collision could prevent loading, especially when combined with the fact that only 10\% of the atoms that transit the FORT were trapped.

\begin{figure}
\includegraphics{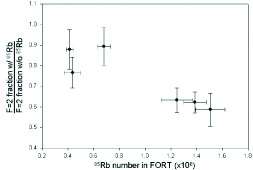}
\caption{\label{label} Change in the fractional amount of $^{87}$Rb in the F=2 ground state as a function of number of $^{85}$Rb loaded into the FORT. At small numbers of $^{85}$Rb in the trap, there is only a small change in the $^{87}$Rb state distribution. However, as the number of $^{85}$Rb increases in the FORT, there is significant change to the ground state distribution of the $^{87}$Rb. Heteronuclear hyperfine-changing and light-assisted collision rates were measures and accounted for in the data presented, and in any case the magnitude of these collision terms was too small to explain the observed changes.} 
\end{figure}

Additional data confirmed that the presence of $^{85}$Rb impacted the optical pumping rates of $^{87}$Rb.  Ground state coherences are expected to suppress the optical pumping from the $^{87}$Rb F=2 to F=1 ground state, and indeed we observe an F=2 to F=1 ground state pumping rate three times slower than would be expected in an unpolarized (and thus decohered) sample.  In a series of experiments, both $^{87}$Rb and $^{85}$Rb were loaded into the FORT, and the change in the steady-state fraction of atoms in the F=2 $^{87}$Rb hyperfine ground state due to $^{85}$Rb atoms was measured. This fractional population was measured by rapidly shutting off both the $^{87}$Rb cooling and repump light during the loading of the FORT and then measuring the F=2 population alone and the total population over several measurements. Figure 4 shows data from such an optical pumping rate experiment.  As the number of $^{85}$Rb in the FORT increases, there is a measurable difference in the ground state distribution of $^{87}$Rb, indicating a change in the ground state coherences and/or m$_{F}$ populations.

In summary, we have observed that the presence of a low density cloud of $^{85}$Rb decreased the load rate of $^{87}$Rb atoms into a shallow FORT. FORT loading is sensitive to ground state coherences, integral to sub-Doppler cooling, thus the reduction in the load rate indicates an induced decoherence in the ground state population. Classical calculations indicate that the dipoles induced in $^{85}$Rb and $^{87}$Rb produce state-dependent forces of sufficient magnitude to account for this decoherence through a process which causes a splitting of the wavepacket similar to a Stern-Gerlach experiment. While this experiment was performed in a thermal gas, with greater control this phenomenon could be used to take advantage of the link between an atom's motion and its internal state. Previous studies have focused on the energy shifts created by induced dipole interactions \cite{Brennen2000}; this work shows that these induced dipoles can significantly alter an atom pair's motion. The presence of these collisions helps to explain a limitation of heteronuclear optical trap loading as well as demonstrating the relevance of induced dipole-dipole collisions in an ultracold gas. 

We would like to acknowledge Drew Davis and Michael DeAngelo for the work they contributed to these experiments. This work was funded by Air Force Office of Scientific Research, grant number FA9550-06-1-0190.


\begin{thebibliography}{31}
\bibitem{density}  This is because for our trap parameters, the MOT peak density is roughly constant; thus a decrease in number corresponds to a decrease in MOT/FORT overlap only along the long axis of the FORT.
\end{thebibliography}
\end{document}